\documentclass[10pt,twocolumn,letterpaper]{article}

\usepackage[pagenumbers]{cvpr}

\usepackage[pdftex]{graphicx}
\usepackage{graphicx}
\usepackage{amsmath}
\usepackage{amssymb}
\usepackage{booktabs}
\usepackage{multirow}
\usepackage[ruled,linesnumbered]{algorithm2e}
\usepackage{siunitx}
\usepackage[accsupp]{axessibility}  
\usepackage[pagebackref,breaklinks,colorlinks]{hyperref}
\usepackage[capitalize]{cleveref}
\crefname{section}{Sec.}{Secs.}
\Crefname{section}{Section}{Sections}
\Crefname{table}{Table}{Tables}
\crefname{table}{Tab.}{Tabs.}

\begin{document}

\title{``Seeing'' Electric Network Frequency from Events}

\author{{Lexuan Xu}$^1$\footnotemark[1], {Guang Hua}$^2$\footnotemark[1], {Haijian Zhang}$^1$\footnotemark[2], {Lei Yu}$^1$\footnotemark[2], {Ning Qiao}$^3$ \\
 $^1${Wuhan University, Wuhan, China} \\
$^2${Institute for Infocomm Research (I$^2$R), A*STAR, Singapore} \\
$^3${SynSense Tech. Co. Ltd., Chengdu, China} \\
{\tt\small \{lexuan.xu,haijian.zhang,ly.wd\}@whu.edu.cn,huag@i2r.a-star.edu.sg,ning.qiao@synsense.ai}}

\maketitle
\renewcommand{\thefootnote}{\fnsymbol{footnote}} 
\footnotetext[1]{Equal contribution. $\dagger$ Corresponding authors. } 
\footnotetext[0]{The research was partially supported by the National Natural Science Foundation of China under Grants 62271354 and the Natural Science Foundation of Hubei Province, China under Grants 2022CFB084 and 2021CFB467.}

\begin{abstract}

Most of the artificial lights fluctuate in response to the grid's alternating current and exhibit subtle variations in terms of both intensity and spectrum, providing the potential to estimate the Electric Network Frequency (ENF) from conventional frame-based videos. Nevertheless, the performance of Video-based ENF (V-ENF) estimation largely relies on the imaging quality and thus may suffer from significant interference caused by non-ideal sampling, motion, and extreme lighting conditions. In this paper, we show that the ENF can be extracted without the above limitations from a new modality provided by the so-called event camera, a neuromorphic sensor that encodes the light intensity variations and asynchronously emits events with extremely high temporal resolution and high dynamic range. Specifically, we first formulate and validate the physical mechanism for the ENF captured in events, and then propose a simple yet robust Event-based ENF (E-ENF) estimation method through mode filtering and harmonic enhancement. Furthermore, we build an Event-Video ENF Dataset (EV-ENFD) that records both events and videos in diverse scenes. Extensive experiments on EV-ENFD demonstrate that our proposed E-ENF method can extract more accurate ENF traces, outperforming the conventional V-ENF by a large margin, especially in challenging environments with object motions and extreme lighting conditions. The code and dataset are available at \url{https://xlx-creater.github.io/E-ENF/}.

\end{abstract}

\begin{figure}[t]
  \centering
   \includegraphics[width=\linewidth]{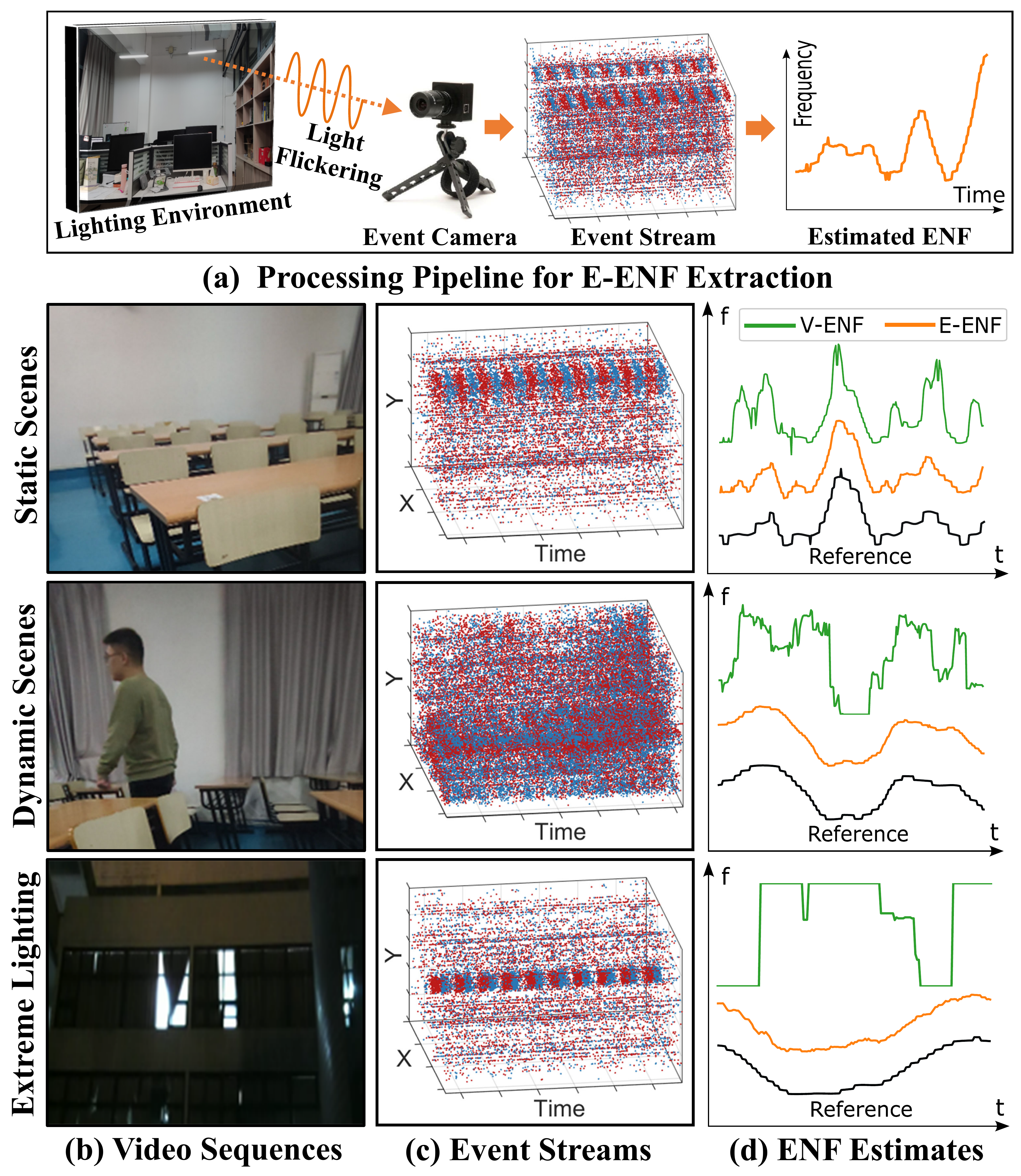}
   \caption{Illustrations of (a) the processing pipeline for E-ENF extraction and comparative experimental results under (b) different recording environments, in terms of (c) recorded event streams and (d) extracted V-ENF and E-ENF.}
   \label{Prototype}
\end{figure}

\section{Introduction}
Light flicker is a nuisance in video~\cite{yang2008effective} but can be exploited to detect and estimate the Electric Network Frequency (ENF)~\cite{garg2011seeing,6553242}, which is the power line transmission frequency in a grid, unlocking the potential of multimedia forensic verification~\cite{garg2021feasibility,frijters2022use}, electrical load monitoring~\cite{sheinin2017computational,shah2019gridinsight}, recording device identification~\cite{vatansever2019analysis,bykhovsky2020recording}, etc. This is enabled by the discovery that the ENF fluctuates slightly and randomly around the nominal value ($50$ or $60$ Hz) and consistently across the entire intra-grid nodes~\cite{GRIGORAS2007136}.

Existing Video-based ENF (V-ENF) extraction methods attempt to restore the illumination fluctuations by averaging the pixel intensities of every frame over the video sequences~\cite{garg2011seeing}. Based on this, V-ENF estimates can be improved by using a higher sample rate in the rolling shutter~\cite{7026086,vatansever2019analysis} or an advanced frequency estimator~\cite{karantaidis2018assessing,choi2019enf,ferrara2021music,han2022phase}. However, due to the inherent characteristics of conventional cameras, existing V-ENF extraction methods still face a series of challenges, which are summarized as follows.

\begin{itemize}
    \item \textbf{Non-ideal sampling.} For global shutter videos, the sample rate equals to the camera's frame rate, usually at $25$ or $30$ fps, which is smaller than the Nyquist rate to record light flicker, resulting in frequency aliasing for ENF estimation~\cite{6553242}. Although rolling shutter can increase the sample rate through the line exposure mechanism, it also introduces the inter-frame idle time within which no signal can be sampled~\cite{vatansever2019analysis,vatansever2022effect}, violating the uniform sampling condition.    
 
    \item \textbf{Motion.} Motion in the video leads to the abrupt change of pixel intensity, which is inconsistent with the flicker pattern and produces strong interference. Note that motion can be caused not only by object movement (\eg, walking person) but also by camera movement (\eg, hand-held camera shake).   

    \item \textbf{Extreme lighting conditions.} Imaging quality of over- or under-exposed videos (even under good lighting conditions) can be substantially degraded, in which the flicker information can be totally lost~\cite{shah2019gridinsight}. Meanwhile, under extreme low-light conditions (\eg, night scene with insufficient lighting), one has to boost the ISO for appropriate exposure, inevitably bringing in ISO noises, which can severely deteriorate the ENF.   
  
\end{itemize}

Generally, the actual content in video recordings can have diverse scenes and objects, which become interference with the task of ENF extraction from the content. Note that putting a video camera still against a white wall illuminated by a light source is an ideal condition for V-ENF extraction~\cite{6553242,7026086}. However, to the very opposite, real-world recordings hardly satisfy this condition. To date, researchers and practitioners are still working intensively to tackle the above problems~\cite{vatansever2017detecting,vatansever2019factors,han2022phase}.

In this paper, we show that the above-mentioned problems can be effectively solved via the proposed use of the so-called event camera, a new modality for recording. The event camera is a neuromorphic sensor that encodes the light intensity variations and asynchronously emits events with extremely high temporal resolution and high dynamic range~\cite{gallego2020event,lichtsteiner2008128}.

Different from V-ENF, the proposed Event-based ENF (E-ENF) extraction approach collects the illumination changes and converts them into an event stream, based on which the ENF traces can be  eventually estimated, as shown in~\cref{Prototype}(a). Thanks to the high temporal resolution and high dynamic range of the event camera, E-ENF can provide a sufficient sample rate and extract reliable ENF traces even under harsh conditions, \eg, motion interference and extreme lighting, as shown in~\cref{Prototype}(b\--d).

Since no prior work has focused on extracting ENF from events, we hereby attempt to provide the first proof-of-concept study theoretically and experimentally demonstrating why the ENF can exist in event streams and how it can be reliably extracted therein, in comparison with the conventional V-ENF approach.

The main contributions of this paper are three-fold:
\begin{itemize}
    \item Based on the event-sensing mechanism, we formulate the process of the ENF fluctuation, reflected by light flicker, being recorded and converted into an event stream, validating the ENF capture in events. 
    \item We propose the first method that effectively extracts the ENF traces from events, featuring a uniform-interval temporal sampling algorithm, a majority-voting spatial sampling algorithm, and a harmonic selection algorithm, respectively.
    \item We construct and open-source the first Event-Video hybrid ENF Dataset (EV-ENFD), containing both events and videos recorded in real-world environments with motion and extreme lighting conditions. The code and dataset are available at \url{https://xlx-creater.github.io/E-ENF/}.
\end{itemize}

\begin{figure*}[t]
  \centering
   \includegraphics[width=.99\linewidth]{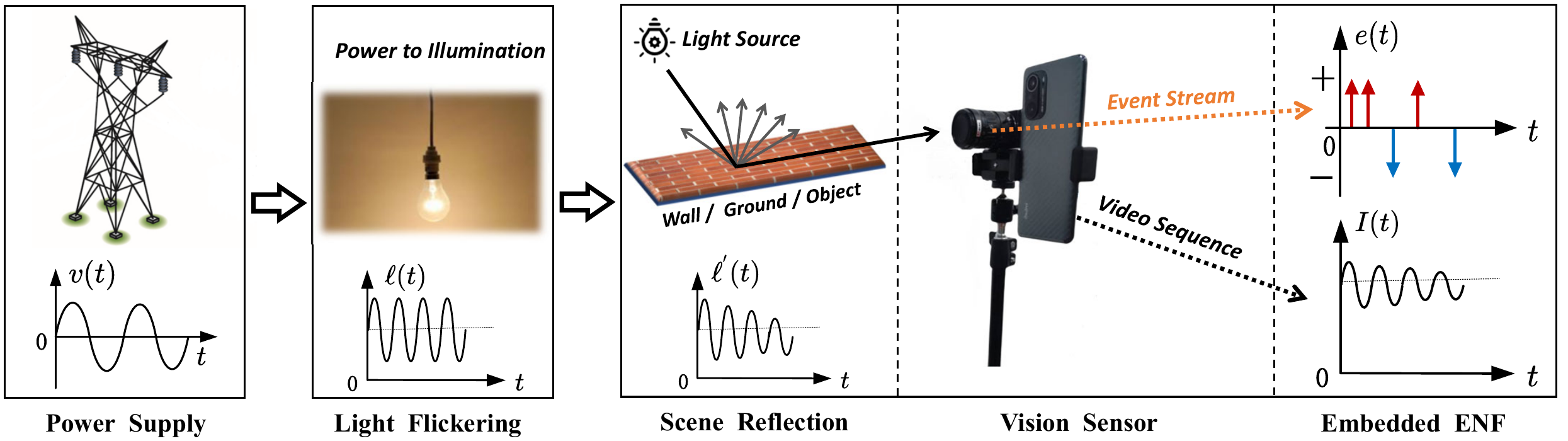}
   \caption{Illustration of ENF capture in video and events. Supplied by voltage $v(t)$ with frequency $\omega_{0}= 2\pi f_{e}(t)$, light illumination $\ell(t)$ varies with its instantaneous power, which fluctuates at $2\omega_{0}$ rad/s. Conventional cameras collect light intensity $\ell^{\prime}(t)$ during the exposure time and convert it to temporal pixel intensity $I(t)$ (see \cite{choi2022invisible} for more about ENF capture in the video). Instead, event cameras perform log transformation on $\ell^{\prime}(t)$ and only generate positive and negative events if the change reaches a certain threshold.}
   \label{ENF}
\end{figure*}

\section{Related Work}
\subsection{Electric Network Frequency}
ENF is the instantaneous power supply frequency in the power grid, whose nominal value varies according to different local standards ($60$ Hz in most of the Americas and $50$ Hz in most rest of the world). Due to the imbalance of instantaneous power supply and consumption, ENF always fluctuates around its nominal value, but this fluctuation is limited to a small range for safety reasons. Under the influence of power supply line differences and the randomness of instantaneous power supply and consumption, the small-scale fluctuations of ENF show obvious regional and temporal differences~\cite{garg2021feasibility}.

Existing research has demonstrated that a light source connected to the grid flickered at exactly twice the ENF, resulting in a light-dark change in illumination that is invisible to the human but can be recorded by cameras~\cite{garg2011seeing,choi2022invisible}, which provides the possibility to extract ENF from videos. The regional and temporal differences of ENF, as well as the characteristics that can be recorded by multimedia files, provide the potential for ENF-based applications, \eg, power monitoring~\cite{sheinin2018rolling}, equipment authentication~\cite{vatansever2019analysis}, and document forensics~\cite{hua2020detection,hua2021robust}.

The principle of the conventional V-ENF estimation method is to extract fluctuation of illumination through pixel intensity changes: according to the different exposure mechanisms~\cite{garg2011seeing,7026086}, the pixel intensity is averaged by frame or by row as the sampling of illumination intensity at the exposure moment. However, due to the inherent characteristics of conventional cameras, this ENF estimation method is easily affected by insufficient camera frame rate, interference from motions, and extreme lighting conditions, resulting in decreased estimation performance.

\subsection{Event Camera}
Unlike conventional cameras, an event camera acts as an asynchronous sensor, each pixel of which responds independently to changes in pixel intensity and generates events to record the changes. The $k$th event, denoted by a tuple $e_{k}=\left(\mathbf{x}_{k} , t_{k}, p_{k}\right)$, is triggered at position $\mathbf{x}_{k} =\left(x_{k}, y_{k}\right)$ and time $t_{k}$, once the log-transformed pixel intensity difference reaches a predefined threshold $C$, \ie,
\begin{equation}
\centering
 \log \big(I\left(\mathbf{x}_{k}, t_{k}\right)\big)-\log \big(I\left(\mathbf{x}_{k}, t_{k}-\Delta t_{k}\right)\big) = p_{k} C,
 \label{event}
\end{equation}
where $I(\cdot)$ represents pixel intensity, $\Delta t_{k}$ is the time difference between the current and the previous event, and the polarity $p_{k} \in\{+1,-1\}$ is the sign of the intensity change\cite{gallego2020event}. The events output by pixel $\mathbf{x}_{k}$ can be represented in the time domain as $e(t) = \sum p_{k} \delta (t-t_{k})$.

Due to this unique form of visual information recording, event cameras have the advantages of high temporal resolution, high dynamic range, and low latency, and have been used in feature detection and tracking~\cite{gehrig2018asynchronous,zhang2021object}, image deblurring~\cite{jiang2020learning,zhang2022unifying}, reconstruction of visual information~\cite{jiang2020robust,liao2022synthetic}, estimation of lighting and its impact removal~\cite{chen2021indoor,wang2022}, and other applications.

\section{Problem Formulation}
\textbf{ENF Modeling.} The grid supply voltage is modeled by
\begin{equation}
\centering
v(t)= V(t) \cos \big (2 \pi f_{e}(t) +\theta \big ) + n(t),
\label{voltage}
\end{equation}
where $V(t)$ and $f_{e}(t)$ are the instantaneous amplitude and frequency, respectively, $\theta$ is the unknown initial phase, and $n(t)$ is the noise. The term $f_{e}(t)$ in \cref{voltage} is the ENF that needs to be estimated, where
\begin{equation}
\centering
f_{e}(t)=  f_0 +  \int_{-\infty}^t f_z(x) dx,
\label{f_e}
\end{equation}
with $f_0$ the nominal value ($50$ or $60$ Hz) and $f_z(t)$ the zero-mean random process characterizing the subtle varying nature of the ENF~\cite{hua2020detection}. The illumination intensity of the grid-connected light source is proportional to the voltage power and fluctuates at the frequency of $2f_{e}(t)$. 

\begin{figure*}[t]
  \centering
   \includegraphics[width=.97\linewidth]{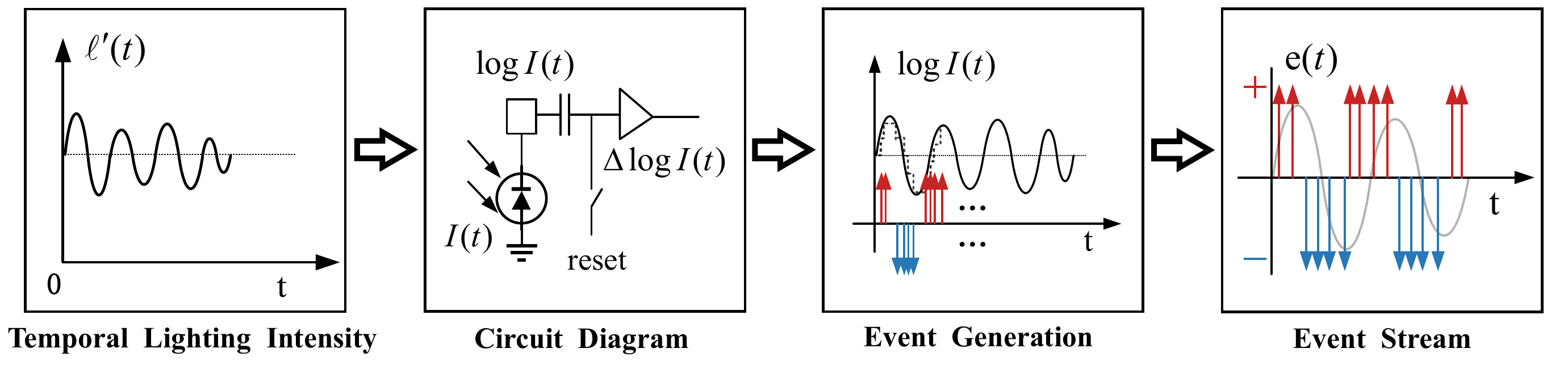}
   \caption{Illustration of the conversion from illumination to events. Each pixel of the event camera independently log-transforms its own pixel intensity $I(t)$, which is then compared with a predefined threshold $C$ to generate positive and negative events.}
   \label{ENF_Event}
\end{figure*}

\textbf{V-ENF Extraction.} The task of V-ENF extraction is to restore the ENF from pixel intensities in the video sequences. As illustrated in~\cref{ENF}, although the amplitude and bias of pixel intensity are significantly different from the original illumination variation, it still preserves the basic frequency information of the ENF. In addition to the ENF-affected illumination change $\ell_\text{ENF}(t)$, there are other components in the pixel intensity $I(t)$, modeled by
\begin{equation}
\centering
I(t) =  G\{\ell_\text{ENF}(t) + \ell_\text{M}(t) + \ell_\text{S} + \ell_\text{N}(t)\},
\label{vidoe problem}
\end{equation}
where $\ell_\text{M}(t)$ and $\ell_\text{S}$ are light differences caused by motion and scene, respectively, and $\ell_\text{N}(t)$ is the noise term. Light intensity is mapped to the pixel intensity $I(t)$ by the camera response function $G\{\cdot\}$. Since the V-ENF approach averages the spatial pixel intensities exposed at the same time, the terms $\ell_\text{M}(t)$ and $\ell_\text{S}$ can cause ENF quality degradation. Besides, in over- or under-exposed videos or under extreme low-light conditions, the function $G\{\cdot\}$ can be severely interfered and lead to incorrect records of ENF-driven illuminations, making it infeasible to extract the ENF.

\textbf{E-ENF Extraction.} For an event camera, an asynchronous event is generated if \cref{event} holds. Within the time period $\tau$, each pixel independently generates events to form the final output of event stream $\mathcal{E}_{}^{\tau}$, given by
\begin{equation}
\centering
\mathcal{E}_{}^{\tau} = \mathcal{E}_\text{ENF}^{\tau}+\mathcal{E}_\text{M}^{\tau}+\mathcal{E}_\text{N}^{\tau},
\label{Event problem}
\end{equation}
where $\mathcal{E}_\text{ENF}^{\tau}$ and $\mathcal{E}_\text{M}^{\tau}$ are events due to illumination changes and motion, respectively, and $\mathcal{E}_\text{N}^{\tau}$ denotes the noisy events. Since the pixels in an event camera are independent, there will be no interference in static scenes. Meanwhile, due to the inherent characteristics of event cameras, the motion of enclosed objects always results in the simultaneous generation of two polarities, unlike illumination events that respond with a single polarity. Therefore, the main problem in E-ENF extraction is to compensate for the polarities caused by motion and suppress noise in event streams.

\section{ENF Extraction from Events}

In this section, we first reveal the connection between the event stream and the ENF of the power grid in \cref{ENF in Event stream}. Then, a simple yet effective event-based ENF extraction algorithm is proposed in \cref{ENF_method}.

\subsection{Modeling ENF with Event Stream}
\label{ENF in Event stream}
Considering a pixel at position $\mathbf{x}_k$ under ideal sinusoidal illumination, we can model its instantaneous intensity~\cite{choi2022invisible} as
\begin{equation}
\centering
I_{\mathbf{x}_{k}}(t)=A(t) \cos \big(4\pi f_{e}(t) +\phi \big)+B(t),
\label{instantaneous intensity}
\end{equation}
where $A(t)$ and $B(t)$ are the instantaneous amplitude and bias, respectively, and $\phi$ is the unknown initial phase. For the event camera, the intensity $I_{\mathbf{x}_{k}}(t)$ is log-transformed and compared with the previous value. An event is emitted as the difference reaches the threshold in \cref{event}, as illustrated in~\cref{ENF_Event}. This mechanism inherently encodes the derivative of log-transformed current $\log \big(I_{\mathbf{x}_{k}}(t)\big)$ as the event emitting rate, \ie, higher derivatives produce more events,
\begin{equation}\label{eq:enf_event}
    |\mathcal{E}^\tau| \propto \frac{d\log \big(I_{\mathbf{x}_{k}}(t)\big)}{dt},
\end{equation}
where $|\cdot|$ denotes the set cardinality and $\tau = [t,t+\Delta t)$ is a short time period starting at $t$. 

Considering log-transformed current $\log \big(I_{\mathbf{x}_{k}}(t)\big)$, we can conduct its relation to the embedded ENF as follows. Let $p\triangleq ({B(t)+\sqrt{B(t)^{2}-A(t)^{2}}})/2$ and $q\triangleq ({B(t)-\sqrt{B(t)^{2}-A(t)^{2}}})/A(t)$, both being functions of $t$, yielding $A=2 p q$ and $B=p(1+q^{2})$, then we have
\[
I_{\mathbf{x}_{k}}(t) 
=p\left(1+q e^{i \omega}\right)\left(1+q e^{-i \omega}\right),
\]
with $\omega = 4\pi f_e(t)+\phi$. So the Taylor expansion of $\log I_{\mathbf{x}_{k}}(t)$ can be neatly expressed,
\begin{align}
\log \big(I_{\mathbf{x}_{k}}(t)\big) &=\log \big(p\left(1+q e^{i \omega}\right)\left(1+q e^{-i \omega}\right) \big) \nonumber \\
&=\log p+\log \left(1+q e^{i \omega}\right)+\log \left(1+q e^{-i \omega}\right) \nonumber\\
&=\log p+2\sum_m(-1)^{m-1} \frac{q^{m}}{m} \cos m \omega,
\label{log transformation}
\end{align}
in which it can be seen that the twice ENF component as well as the harmonics are preserved by the summation. 
Since the bias $B$ is always greater than the amplitude $A$ (see sine waves in \cref{ENF,ENF_Event}), it then follows that $p>0$ and $|q|<1$, satisfying the condition for Taylor approximation.

Based on \cref{eq:enf_event} and \cref{log transformation}, the ENF is embedded in the event stream and can be extracted directly from $\mathcal{E}^\tau$. Moreover, the original time-varying amplitude and bias differences are suppressed after the logarithm, which can be beneficial to the subsequent ENF estimation.

\begin{figure*}[t]
  \centering
   \includegraphics[width=.96\linewidth]{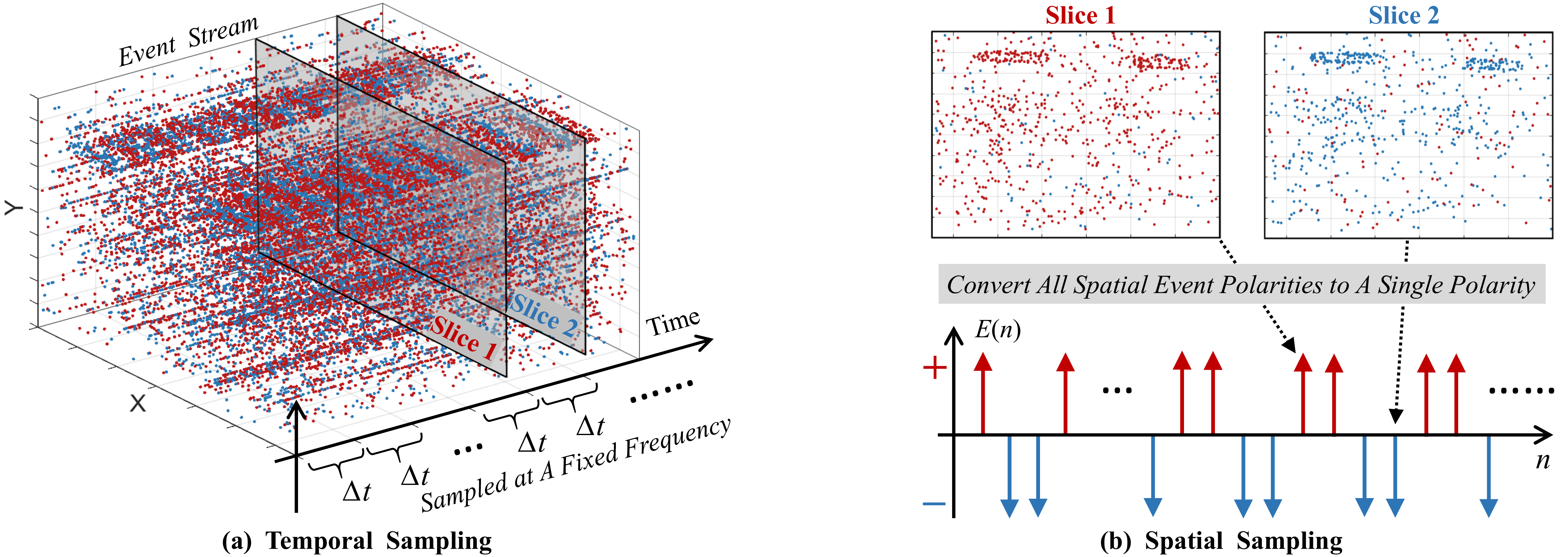}
   \caption{Illustration of generating uniformly discrete events. (a) Uniform-interval temporal sampling. It reserves events within a selected time interval. (b) Majority-voting spatial sampling. For each sample, it converts all spatial event polarities into a single polarity result.}
   \label{method}
\end{figure*}


\subsection{Event-based ENF Estimation}
\label{ENF_method}
The independent pixels in the event sensor can generate highly dynamic and extremely rich events containing the ENF information, but the event camera works asynchronously and the output time-spatial binary events, as depicted in \cref{method} (a), cannot be treated as a conventional sampled signal for further processing due to the high dimension and non-uniform sampling. Therefore, we propose the following treatments to solve this problem and yield accurate ENF estimates.

\subsubsection{Uniform-Interval Temporal Sampling}
The event-sensing mechanism in \cref{event} indicates that the recorded events are non-uniform in both temporal and spatial domains. In this subsection, we propose a uniform-interval temporal sampling mechanism to frame the asynchronous events into uniform time intervals. Let the desired sampling interval be $\Delta t$, then at the $n$th time slice, the corresponding spatial events are recorded as the sampled raw data at time index $n$, \ie, $\tilde{E}(n)$, while the remaining events within the interval are discarded. Since it is possible that at certain time slices there is no event recorded, to avoid missed sampling, we propose to use the events in the nearest future as the sampled data. Thanks to the high sensitivity of the event-sensing mechanism, a reasonable choice of $\Delta t$ can ensure sufficient data sampling and high ENF sampling frequency at the same time. The above uniform sampling process is expressed as
\begin{equation}
  \tilde E(n) = \left\{ {{e_k}\left| {\arg \mathop {\min }\limits_k \;{\kern 1pt} {t_k} \geqslant {t_1} + n\Delta t} \right.} \right\},
  \label{time sampling}
\end{equation}
where $t_{1}$ is the time index of the first event slice. The above process is summarized
in~\cref{Uniform Temporal Sampling}.

\begin{algorithm}[!t]
    \caption{Temporal Sampling}
    \label{Uniform Temporal Sampling} 
    \KwIn{event stream $\mathcal{E}_{}^{\tau}$, sampling interval $\Delta t$\;}
    \KwOut{slices of events $\tilde{E}(n)$\;}
    $t_{1} \gets \text{start time of event stream} \ \mathcal{E}_{}^{\tau}$\;
    $t_{N} \gets \text{end time of event stream} \ \mathcal{E}_{}^{\tau}$\;
    Initialize {sampling moment $ t_{n} \gets t_{1}$\;}
    \While{$t_{n} < t_{N}$}{
       $\tilde{E}(n)$ $\gets$ set of events in $\mathcal{E}_{}^{\tau}$ with timestamp $t_{n}$\;
       \If{$\tilde{E}(n)${\rm{ is empty}}}
          {$\tilde{E}(n) \gets$ events in the nearest future of $t_{n}$\;} 
        $t_{n+1} \gets t_{n} +\Delta t$\;
        $n \gets n+1$\;
    }
    Return $\tilde{E}(n)$\;
\end{algorithm}

\subsubsection{Majority-Voting Spatial Sampling}
After uniform temporal sampling, we obtain the event slices at each sampling moment.
Since the independent pixels correspond to different spatial content of the scene and respond to illumination differently, the spatial distribution of events can vary substantially from slice to slice. Besides, there are illumination-independent events at each slice due to noise and motion, and the number of these events increases when the motion is intense. Under the influence of these factors, the event polarity and location are usually inconsistent, as shown in \cref{method} (b) upper plots, making it difficult to judge the change of illumination. 

Fortunately, the motion produces bipolar events while the illumination produces unipolar events at each sampling moment, which provides the possibility for us to judge the illumination polarity. Inspired by this foundation and the principle of \textit{Law of Large Numbers}, we treat each event in the slice as an independent sampling of intensity change on a majority-voting basis, \ie, in each slice, we consider the polarity occurring more frequently as the corresponding illumination polarity, as shown in \cref{method} (b) lower plot.

The spatial sampling process can be expressed as the comparison of the number of positive and negative events at each sampling time, which can be normalized by the sign function $\mathrm{sgn}(\cdot)$, \ie,
\begin{equation}
  E(n) = \mathrm{sgn}(N^n_{\text{pos}}-N^n_{\text{neg}}),
\end{equation}
where $N^n_{\text{pos}}$ and $N^n_{\text{neg}}$ are the number of positive and negative events at the $n$th sampling moment, respectively.

After temporal and spatial sampling of the event stream shown in \cref{method}, we obtain a one-dimensional polar sequence $E(n)$ recorded at a selected sampling interval $\Delta t$, which encodes illumination changes in standard event polarity instead of absolute pixel intensities.

 \subsubsection{Harmonic Selection}
 According to \cref{log transformation}, the ENF exists in the harmonic form including both the second and higher-order ones. Therefore, in addition to bandpass filtering the polarity sequence $E(n)$ at twice the ENF nominal value, we can also filter it at higher harmonic components to exploit more useful information. The filtered sequences are then analyzed at each harmonic using a Short-Time Fourier Transform (STFT) and peak search to obtain the corresponding dominant time-frequency variation at each harmonic component. Although ENF-related time-frequency traces can be obtained at each harmonic, the results at some harmonics are inevitably corrupted. In this case, we propose a harmonic selection mechanism to ensure that all the considered harmonics can contribute instead of interfering with the time-frequency estimation process.

 \begin{algorithm}[!t]
    \caption{Harmonic Selection}
    \label{Harmonic Selection}
    \KwIn{Binary polarity sequence $E(n)$, to-be-estimated harmonic order $M$, time window length $T$\;}
    \KwOut{ENF estimates $f_{E}$\;}
    Time-frequency trace at the $m$th harmonic $f_{m}$ $\gets$ normalize the STFT and peak search results of $E(n)$ at first $M$ harmonics to baseband, $m\le M$\;
    $m$th harmonic $f_{m} = \sum\limits_{l}^{}  f_{m}[lT:(l+1)T]$ $\gets$ divide harmonic results into time segments by $T$\;
    \For{{\rm each time segment}}{
    Calculate the smoothness score $S$ using~\cref{S}\;
    $h$ $\gets$ harmonic index of the smallest $S$\;
    $f_{E}[lT:(l+1)T]$ $\gets$ $f_{h}[lT:(l+1)T]$\;}
    Return $f_{E}$\; 
\end{algorithm}

\begin{figure}[!t]
  \centering
   \includegraphics[width=0.98\linewidth]{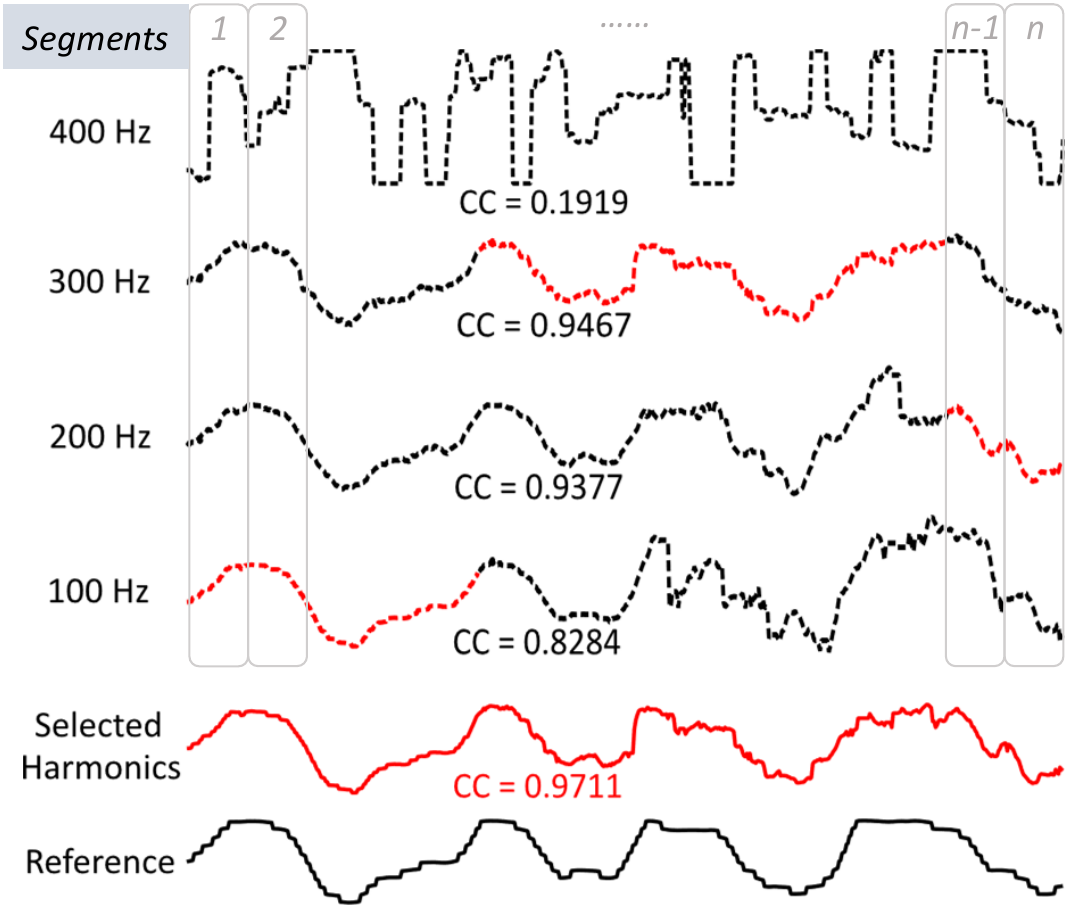}
   \caption{An example of harmonic selection. In this scene, the event camera shoots at a fan that is stationary for the first half and rotates for the second half. During the shooting, most pixels of the  camera generate events by the rotation of the fan. The harmonic selection result using~\cref{Harmonic Selection} is further improved compared with the estimation results using each harmonic individually.}
   \label{ENF_Har}
\end{figure}

Inspired by the work in~\cite{vatansever2022enf} which analyzes the variation of ENF estimation results and proposes to only retain the part that conforms to the ENF fluctuation, we normalize each harmonic estimation result to the baseband and divide each result into segments with respect to time. In each segment, the harmonic result that most conform to the tiny and slowly varying nature of ENF is selected as the estimation result of the corresponding time. The smoother the variation, the closer it is to the fluctuation of the real ENF~\cite{hua2021robust}. Mathematically, we use the sum-absolute-first-order difference function $S$ to characterize the smoothness, given by 
\begin{equation} 
 S_{m} = \sum_{n=1}^{N_{f}}\left|f_{m}[n]-f_{m}[n-1]\right|,
 \label{S}
\end{equation}
 where $m$ is the harmonic index, $N_{f}$ represents the number of points of ENF estimates within the $l$th segment. The harmonic selection process is summarized in~\cref{Harmonic Selection} according to the above mechanism, and the smoothest result in each time segment is selected as the final estimation.  Harmonic selection can lead to better ENF estimates in the presence of a large number of illumination-independent events. A demonstrative example is provided in \cref{ENF_Har}, in which the estimates based on the proposed harmonic selection are the closest to the ground-truth reference.

\section{Experiments}

\begin{figure*}[t]
  \centering
   \includegraphics[width=.98\linewidth]{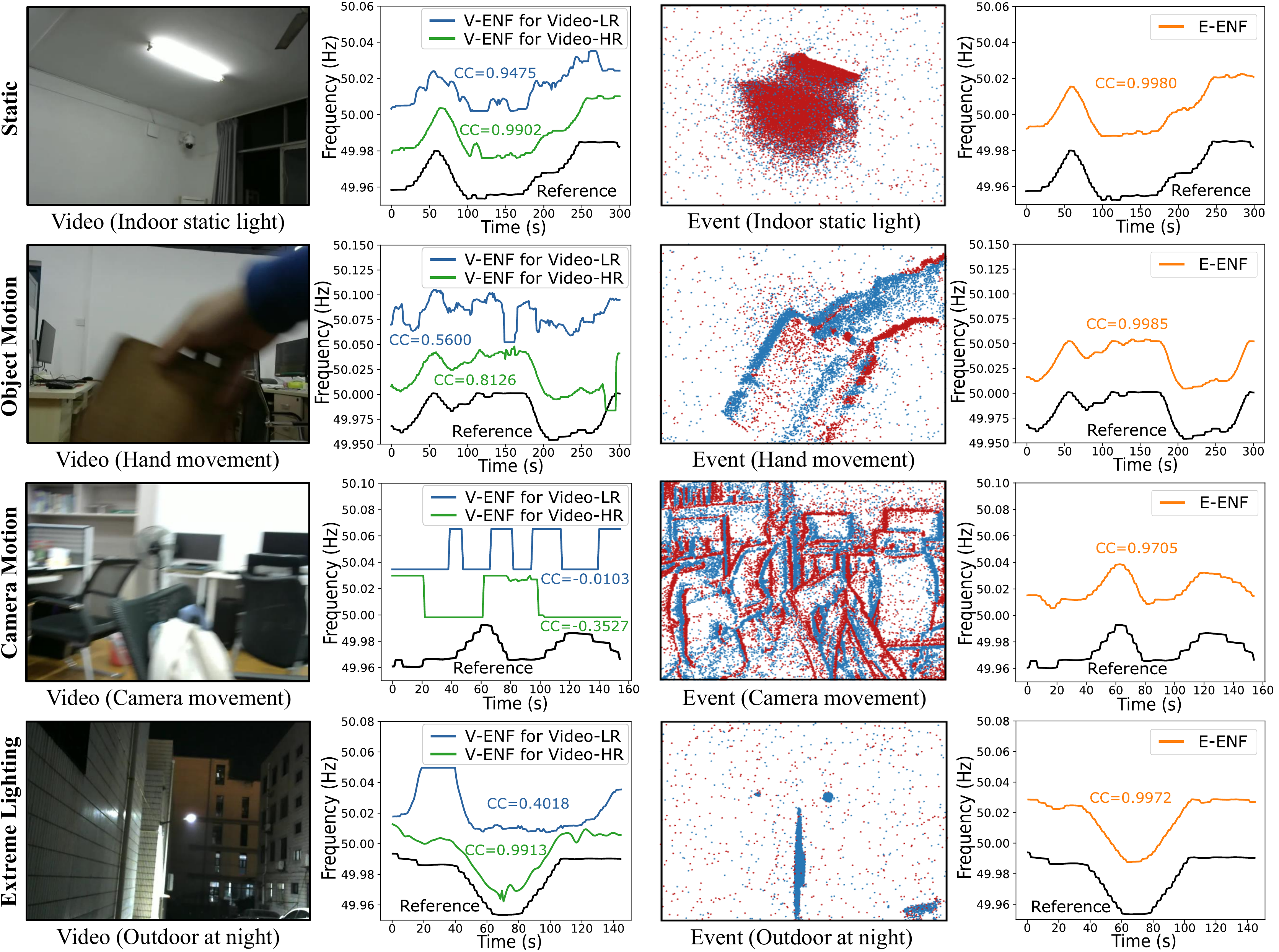}
   \caption{Comparisons of experimental results between V-ENF and the proposed E-ENF under different environments and sequences (in~\cref{tool}). \textbf{First row:} An indoor static scene with strong illumination. \textbf{Second row:}  A dynamic scene with object movement. \textbf{Third row:}  A dynamic scene with camera movement. \textbf{Fourth row:}  An outdoor extreme lighting scene with weak illumination at night. Note that different biases and frequency scaling are added to the estimated ENF curves for better visual quality.}
   \label{results}
\end{figure*}

\subsection{Dataset and Experimental Settings}
\textbf{Event-Video hybrid ENF Dataset (EV-ENFD).} Since the characteristics of the rolling shutter provide a high sample rate mitigating the aliasing problem at the frame level~\cite{6553242}, we construct the EV-ENFD for real illumination scenes using two cameras with a rolling shutter mechanism to acquire videos and an event camera to acquire events, and details on capturing instruments are shown in~\cref{tool}. All cameras are fixed on a tripod to ensure synchronized recording time and scene. We manually adjusted the shutter speed and ISO of Redmi K40 to $1/200$ \SI{}{\second} and $1600$, respectively, to enable the perceptible ENF traces in videos. In addition, due to the built-in stabilizer in Redmi K40, the recorded content is less affected by hand-held camera shake than the other two cameras.

\setlength{\tabcolsep}{1.4pt}{
\begin{table}
  \caption{The various cameras used in the experiments and their respective resolution when used. Note that the shutter speed and ISO of the Video-HR are artificially adjusted to $1/200$ \SI{}{\second} and $1600$ to ensure the perceptible ENF traces in the videos.}
  \label{tool}
  \centering
\small
\setlength\tabcolsep{8pt}
  \begin{tabular}{ l  l  l }
  \hline
  \textbf{Seq.} & \textbf{Cameras} & \textbf{Spatial Resolution}\\
  \hline
  Events  & IniVation DVXplorer 
  & $640\times480$ \\
  Video-LR & Logitech C270 & $640\times480$\\ 
  Video-HR & Redmi K40  & $1920\times1080$\\
  \hline
  \end{tabular}
  \end{table} }

  \begin{table*}[ht]
\caption{Quantitative evaluations of the V-ENF and the proposed E-ENF applied to the recordings in the event-video ENF dataset. Average CCs and MAEs are used to evaluate the similarity and absolute error between estimated ENFs and their truth references.}
\label{table2}
\centering
\small
\setlength\tabcolsep{11.2pt}
\begin{tabular}{cclcccccc}
\hline
\multirow{2}{*}{\bf Sequences}    & \multicolumn{2}{c}{ \multirow{2}{*}{\bf Method}} 
   & \multicolumn{2}{c}{\bf Static Scenes}     
   & \multicolumn{2}{c}{\bf Dynamic Scenes}    
   & \multicolumn{2}{c}{\bf Extreme Lighting}  \\ \cline{4-9} 
   & \multicolumn{2}{c}{}   
   & \multicolumn{1}{c}{\textbf{CC}} & \textbf{MAE}    
   & \multicolumn{1}{c}{\textbf{CC}} & \textbf{MAE}    
   & \multicolumn{1}{c}{\textbf{CC}} & \textbf{MAE}    \\ \hline
Video-LR           & \multicolumn{2}{c}{V-ENF\cite{7026086}}      
& \multicolumn{1}{c}{$0.6434$}    & {$13.7 \times 10^{-3}$} 
& \multicolumn{1}{c}{$0.3970$}    & {$24.2 \times 10^{-3}$} 
& \multicolumn{1}{c}{$0.0557$}    & {$32.7 \times 10^{-3}$} \\ 
Video-HR           & \multicolumn{2}{c}{V-ENF\cite{7026086}}      
& \multicolumn{1}{c}{\underline{$0.8906$}}  & {\underline{$6.2 \times 10^{-3}$}} 
& \multicolumn{1}{c}{\underline{$0.7538$}}  & {\underline{$11.1 \times 10^{-3}$}} 
& \multicolumn{1}{c}{\underline{$0.7602$}}  & {\underline{$10.1 \times 10^{-3}$}}\\ \hline
Events             & \multicolumn{2}{c}{E-ENF}                   
& \multicolumn{1}{c}{\textbf{$\bf 0.9862$}} & {\textbf{$\bf 1.9 \times 10^{-3}$}} 
& \multicolumn{1}{c}{\textbf{$\bf 0.9553$}} & {\textbf{$\bf 2.1 \times 10^{-3}$}} 
& \multicolumn{1}{c}{\textbf{$\bf 0.9867$}} & {\textbf{$\bf 2.0 \times 10^{-3}$}} \\ \hline
\end{tabular}
\vspace{-0.6em}
\end{table*}

The EV-ENFD has three categories according to the capturing conditions: {\it static} for static scenes without relative motions, {\it dynamic} for dynamic scenes caused by object or camera motions, and {\it extreme lighting} for the scenes with over or under-exposed regions. In summary, the EV-ENFD contains a total of $51$ sets of data composed of $16$ static scene data, $25$ dynamic scene data, and $10$ extreme lighting scene data. Each set consists of a stream of events, a low-resolution video recorded with low ISO (Video-LR in~\cref{tool}), and a high-resolution video recorded with high ISO (Video-HR in~\cref{tool}). 
 
To examine and evaluate the quality of the extracted V-ENF and E-ENF traces, the EV-ENFD also includes the corresponding ground-truth ENF reference for each set. We use a computer sound card to directly record the grid voltage changes after a step-down transformer, according to the works in~\cite{Cooper2010AnAA,pop2017fast}. An STFT with a 16-second time window and 1-second step (overlapped by $15$ seconds) is used to obtain the ENF reference.

\textbf{Experimental Settings.} The V-ENF estimation method proposed in~\cite{7026086} is employed for the estimation of ENF in videos. Considering the dynamic nature of the scenes encountered by the EV-ENFD, we only average two consecutive frames to suppress background content instead of averaging all frames to reduce the interference caused by motions. By calculating the average pixel intensity, the extracted visual information is subtracted to eliminate the scene and motion effects as much as possible. For ENF estimation in events, we choose the sampling interval $\Delta t =$ \SI{0.001}{\second} in the process of uniform-interval temporal sampling and perform harmonic selection with a time window length of \SI{10}{\second}. The same STFT setting as in ENF reference is used in the V-ENF and the proposed E-ENF method.

\subsection{Qualitative Analysis}
The V-ENF and E-ENF extraction results under several representative recording conditions are shown in~\cref{results}. It can be seen that for low-resolution and low-ISO videos, \ie, Video-LR, the estimation results of V-ENF have ineligible noise even in a completely static scene with sufficient illumination. In dynamic and extreme lighting scenes, the performance of the V-ENF drops dramatically than in static scenes, resulting in ENF estimates that do not match the ground-truth reference. For high-resolution and high-ISO videos, \ie, Video-HR, the performance of the V-ENF can be improved, because a higher ISO can increase the sensitivity to light flicker and obtain more samples. Therefore, in static and extreme lighting scenes, the V-ENF estimation method achieves relatively satisfactory ENF estimates. However, in dynamic scenes, especially the motion caused by the hand-held camera, a large number of illumination-independent intensity changes appear, which seriously degrades the performance of the V-ENF approach. 

In contrast, the proposed E-ENF extraction method not only has stable performance in static scenes but more importantly also shows advantages under the conditions of motion interference and extreme lighting. The motion of enclosed objects produces nearly the same number of positive and negative events, while the illumination change produces a single polarity event. The remaining influence of motion after positive and negative polarity cancellation of enclosed objects is well eliminated by the proposed majority-voting spatial sampling mechanism. This is why the proposed E-ENF approach can still maintain promising performance under different motion conditions. In addition, since event cameras have inherently high temporal resolution and high dynamic range, they are more sensitive to illumination changes compared to conventional cameras, which allows the E-ENF to capture the ENF even under extreme lighting conditions. For all the above reasons, the E-ENF approach achieves the best ENF estimates across diverse real-world illumination scenes. 

\subsection{Quantitative Analysis}
The quantitative experimental results are summarized in \cref{table2}, in which the \textit{Pearson} Correlation Coefficient (CC) and Mean Absolute Error (MAE) between estimated results and their ground truth are used as the metrics. We make the following observations from this table. First, due to more perceptible ENF traces in videos, the V-ENF can obtain better estimation results for Video-HR than those for Video-LR, especially in extreme lighting conditions. A high ISO makes it possible to capture subtle illumination changes in videos, thus the performance of CC and MAE values are substantially improved compared to those achieved by a low ISO. Meanwhile, for static and dynamic scenes, it is further seen that the V-ENF approach cannot effectively solve the problem of illumination-independent intensity interference, even with scene and motion cancellation in advance.

The proposed E-ENF approach has achieved improved performance across all settings by clear margins. Specifically, for the three types of scenes in the EV-ENFD, the E-ENF approach can effectively record illumination changes, consistently capturing the ENF traces. For the metric of CC, the similarity between the estimated ENF traces and the ground-truth reference is greater than $0.95$, which significantly exceeds the performance of the V-ENF. For the metric of MAE, we can see a larger performance margin between the V-ENF and our proposed E-ENF, especially in dynamic and extreme lighting scenes. It is worth mentioning that our E-ENF is conducted without performing complex scene and motion removal operations.

\section{Conclusion}

In this paper, we have proposed a novel ENF extraction approach from the so-called events (E-ENF), a new modality of recording. Taking advantage of the high temporal resolution and high dynamic range of the event camera, our method overcomes the shortcoming of non-ideal sampling in the existing V-ENF estimation approaches and also obtains promising ENF estimates under motion and extreme lighting conditions that V-ENF methods struggle to handle. The proposed approach greatly expands the applicability of the ENF and provides a new direction for related research and application. To validate and measure the performance of ENF estimation from this new modality and by conventional means, we have constructed an event-video hybrid ENF dataset termed EV-ENFD, containing both static scenes and scenes with motion and under extreme lighting conditions. Experimental results based on the EV-ENFD confirm the advantage of the proposed approach. 

{\bf Limitations.} In contrast to V-ENF, the scope of applicability for E-ENF is limited due to the relatively infrequent use of event cameras. Furthermore, the sluggish generation of events in the presence of weak light flickering can result in  missing crucial illumination information, which ultimately leads to the deterioration of E-ENF performance.

{\small
\bibliographystyle{ieee_fullname}
\bibliography{egbib}
}

\end{document}